\documentclass[%
 aip,
 amsmath,amssymb,
 reprint,%
]{revtex4-1}
\usepackage{graphicx}
\usepackage{dcolumn}% Align table columns on decimal point
\usepackage{bm}% bold math
\usepackage[utf8]{inputenc}
\usepackage[T1]{fontenc}
\usepackage{braket}
\usepackage{mathptmx}
\usepackage{etoolbox}
\usepackage{color}
\usepackage{amssymb}
\usepackage{amsthm}
\usepackage{dirtytalk}
\usepackage{textcomp}
\usepackage{gensymb}
\usepackage{booktabs}
\usepackage{subfigure}
\usepackage{mathtools}
\usepackage{threeparttable}
\usepackage{gensymb}

\usepackage[utf8]{inputenc}
\usepackage[T1]{fontenc}
\usepackage{mathptmx}
\usepackage{etoolbox}

\makeatletter
\def\@email#1#2{%
 \endgroup
 \patchcmd{\titleblock@produce}
  {\frontmatter@RRAPformat}
  {\frontmatter@RRAPformat{\produce@RRAP{*#1\href{mailto:#2}{#2}}}\frontmatter@RRAPformat}
  {}{}
}%

\makeatother
\begin{document}
\preprint{AIP/123-QED}

%\title{\textcolor{black}{Modeling energy transfer properties and absorption spectra in layered metal-organic frameworks based on a Frenkel-Holstein Hamiltonian}}
\title{Frenkel/charge transfer Holstein Hamiltonian applied to energy transfer in 2D layered metal-organic frameworks}

\author{David Dell'Angelo$^*$}
\affiliation{Department of Chemistry and Environmental Science, New Jersey Institute of Technology, Newark 07102, NJ United States}
\author{Mohammad R. Momeni}
\author{Shaina Pearson}
\author{Farnaz A. Shakib$^*$}
\affiliation{Department of Chemistry and Environmental Science, New Jersey Institute of Technology, Newark 07102, NJ United States}
\email{dd64@njit.edu,shakib@njit.edu}

\begin{abstract}
Optimizing energy and charge \textcolor{black}{transfer} is key in design and implementation of efficient layered conductive metal-organic frameworks (MOFs) for practical applications. In this work, for the first time, we investigate the role of both long-range excitonic and short-range charge transfer coupling as well as their dependency on reorganization energy on through-space \textcolor{black}{charge transfer} in layered MOFs.
A $\pi-$stacked model system is built based on the archetypal Ni$_3$(HITP)$_2$, HITP = 2,3,6,7,10,11-hexaiminotriphenylene, layered MOF and a Frenkel/charge transfer Holstein Hamiltonian is developed that takes into account both electronic coupling and intramolecular vibrations. \textcolor{black}{The dependency of the long and short-range couplings of secondary building units (SBUs) and organic linkers to stacking geometry are evaluated which predicts that photophysical properties of layered MOFs critically depend on the degree of ordering between layers.} We show that the impact of the two coupling sources in these materials can be discerned or enhanced by displacement of the SBUs along the long- or short molecular axes. The effects of vibronic spectral signatures are examined in both perturbative and resonance regimes. Although, to the best of our knowledge, displacement engineering in layered MOFs currently remains beyond reach, the findings reported here offer new details on the photophysical structure-property relationships in layered MOFs and provide suggestions on how to combine elements of molecular design and engineering to achieve desirable properties and functions for nano- and mesoscale optoelectronic applications.

\end{abstract}
\maketitle
\begin{figure}[!h]
%\centering
\includegraphics[width=0.99\linewidth]{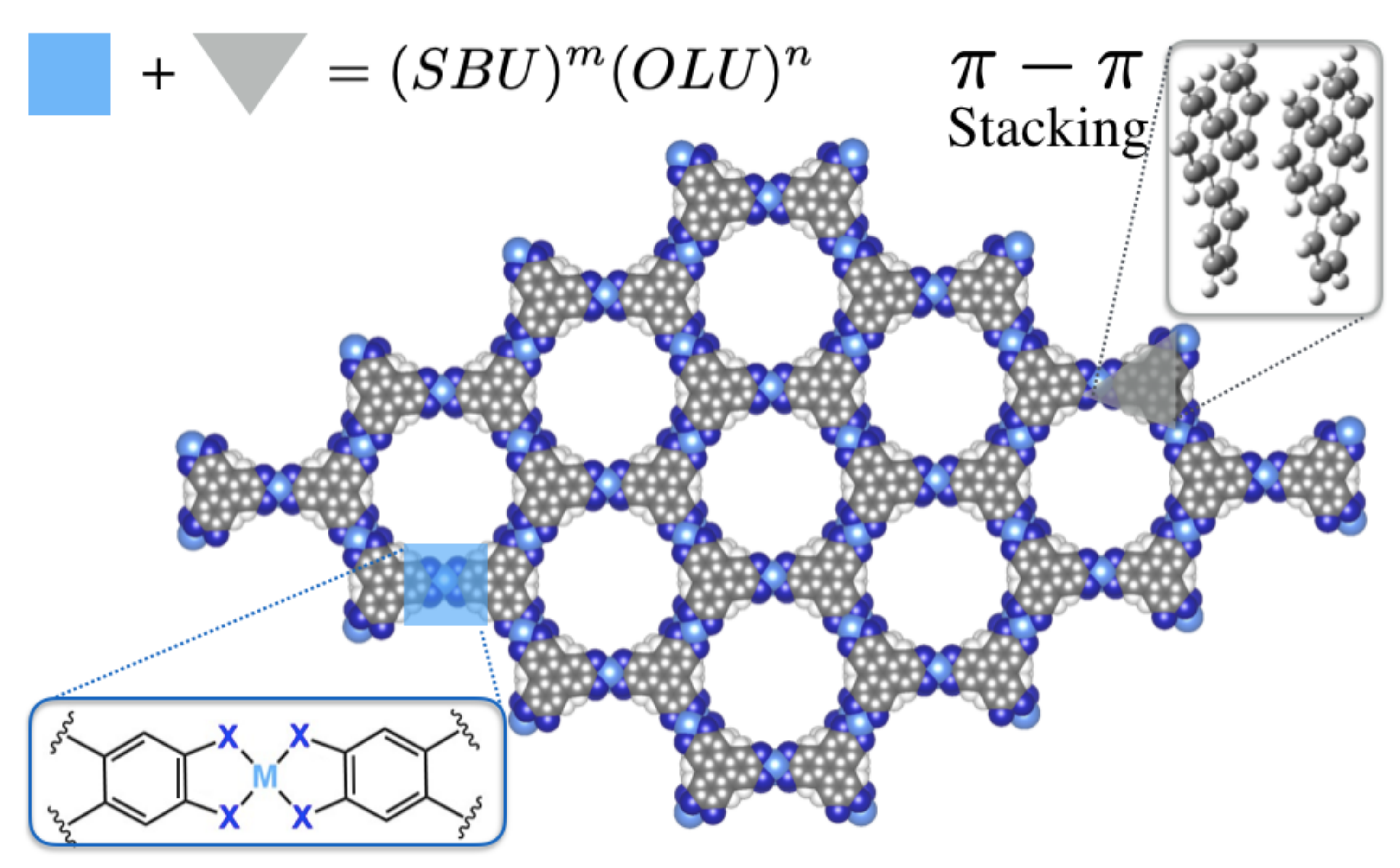}
\caption{Representative layered architecture of a conductive $\pi-$stacked MOF highlighting the ``secondary building unit (SBU, blue square) and organic linker unit (OLU, gray triangle)" belonging to the family of layered MOFs based on triphenylene linkers with square planar metal centers.\cite{camp15} The inset figures highlight SBU at the bottom left and $\pi$-stacked linkers on the top right.}
\label{fig:1}      
\end{figure}

\noindent\textbf{\large I. Introduction\label{Intro}}
\\~\\
In recent years, $\pi-$stacked metal--organic frameworks (MOFs) have emerged as a new class of layered materials which offer record-breaking electrical conductivity along with permanent porosity and high surface area.\cite{Chakraborty21,xie-sko20,ko18} As depicted in Figure~\ref{fig:1}, layers of $\pi-$stacked MOFs are generally built from tetra--coordinated metal nodes and electron-rich $\pi-$conjugated organic linkers extended in the \textit{ab} plane. The bulk architecture is formed from stacking of these layers along the \textit{c} direction via van der Waals interactions. 
The highly ordered and tunable structure of layered MOFs by systematic substitution of metal nodes and organic linkers, which also allows compact device implementation, promotes their applications not only as electrochemical but also photochemical energy conversion and storage systems.\cite{xie-sko20,qiu20,nyak20} Further progress in utilizing layered MOFs for such applications requires a deeper understanding of the structure--function relationships. Fortunately, the well-defined modular structure of layered MOFs and the possibility of linker--based luminescence provides a means to correlate structure and composition to photon capture, as has been extensively probed for conventional 3D MOFs.\cite{allen09} However, understanding charge and energy transport in layered MOFs can be much more complicated~\cite{xie-sko20} than their 3D counterparts due to the possibility of not only through--bond but also through--space charge and energy transport mechanisms.\cite{xie-sko20} Both of these pathways can be drastically affected by intrinsic dynamical motions of the layers.\cite{Shi:2020,Momeni:2021} We have recently demonstrated that interlayer displacements and slipping of the $\pi-$stacked layers compared to each other 
may affect the metal--to--semiconductor transition in layered MOFs.\cite{zeyu21} This is similar to observations in $\pi$--stacked molecular aggregates where inter--molecular displacements as small as 1.6 \AA~generate a dramatic impact on photophysical properties.\cite{gregg08} Such observations signify the importance of evaluating the correlations between
interlayer displacements and photophysical properties in $\pi$--stacked layered MOFs. 

To study \textcolor{black}{charge transfer}, one may employ the Frenkel exciton model~\cite{frenkel31} which is commonly used to investigate $\pi-$stacked molecular aggregates.\cite{spano18} This model relates absorption/emission spectral features of aggregates to electronic properties of monomers and the interaction between individual electronically excited states which include both long--range and short--range interactions. 
Indeed, there are very recent experimental studies where photophysical behaviours of layered MOFs, or layered MOFs included heterostructures, were analyzed with a focus on either long--range excitonic coupling ($J$)~\cite{cao17,liao18} 
or short--range charge transfer coupling ($T$).\cite{liang19,nyak20} However, in reality, the stacked architecture of layered MOFs allows for both short-- and long--range interactions to be simultaneously present, similar to the situation in $\pi-$stacked 
molecular aggregates.\cite{spano15} Hence, instead of considering \textit{J} and \textit{T} separately, they have to be considered on equal footings.\cite{camp15} Most importantly, they should scale with reorganization energy ($\lambda$),\cite{ogih17} defined as the energy changes associated with geometry relaxations during charge/energy transfer. This aspect brings up the question of how the $J/\lambda$ and $T/\lambda$ ratios modulate the through--space~\cite{ko18} energy \textcolor{black}{transfer} (ET)\footnote{Throughout this manuscript, ET refers to both charge and energy \textcolor{black}{transfer} along the $\pi-\pi$ stacking direction.} along the $\pi-\pi$ stacking direction of layered MOFs.

\begin{figure}[ht]
  \centering
\includegraphics[width=0.99\linewidth]{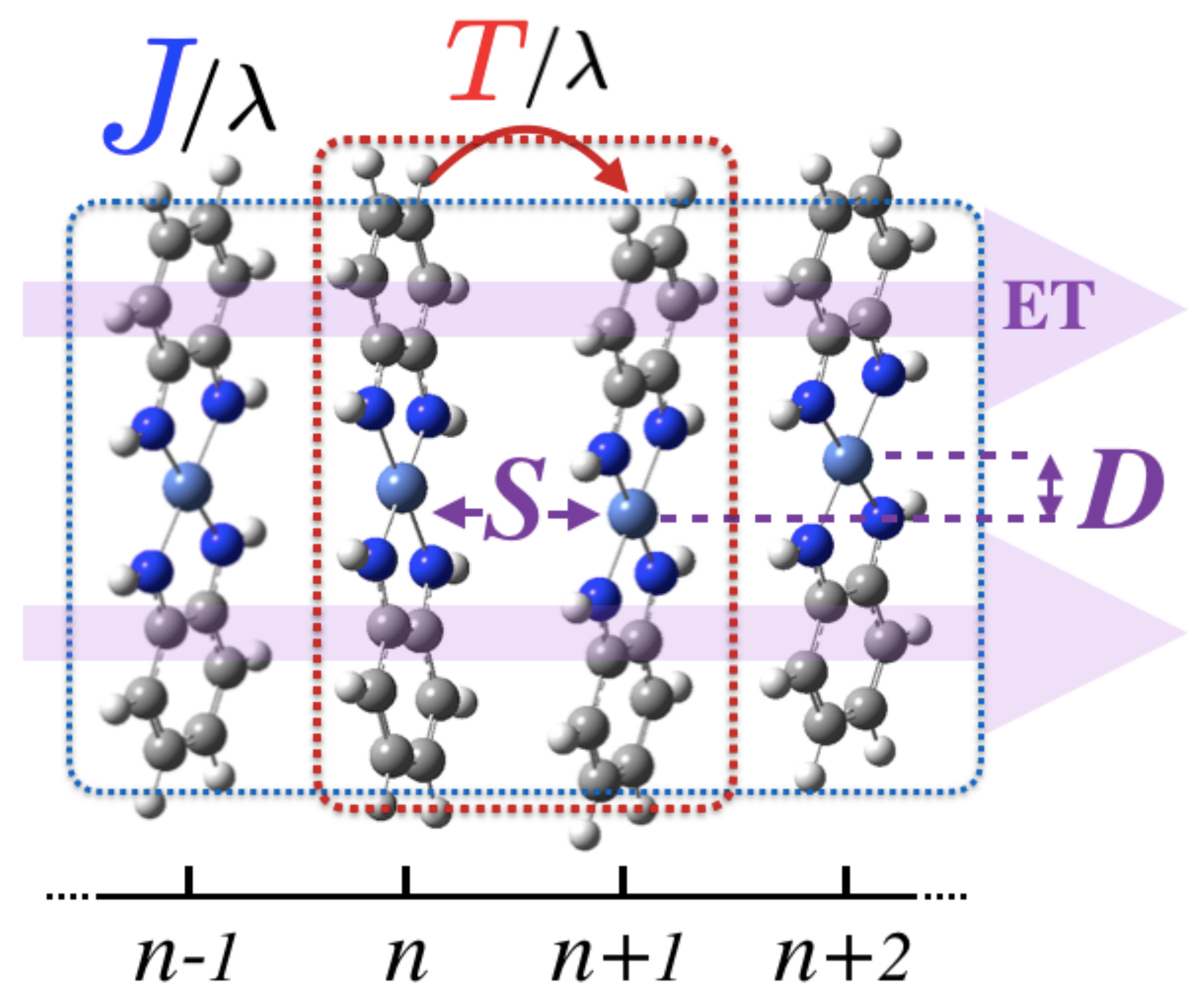}
\caption{The model used to study the ET process along the $\pi-\pi$ stacking direction in layered MOFs. Blue and red dashed frames represent the long and short range interactions determined by the $J$ and $T$ parameters, respectively. In turn, these quantities scale with the reorganization energy $\lambda$ and depend on the interlayer spacing ($S$) and displacement ($D$) between consecutive units.}
\label{fig:2}
\end{figure}

To answer such questions, in this work, we model through--space ET in layered MOFs by projecting $J$ and $T$ onto the organic linker and inorganic secondary building block subunits of the material, shortened as OLU and SBU, where the latter corresponds to a square planar metal center bonded to two adjacent organic linkers (see Figure ~\ref{fig:1}).
$J$ and $T$ are then evaluated according to the inter--molecular displacements and rotations. 
Given its high performing conductivity,\cite{hendon19} we model these parameters on the archetypal 
layered MOF~\cite{shebe14} Ni$_3$(HITP)$_2$, HITP=2,3,6,7,10,11--hexaiminotriphenylene.\cite{shebe14,day19,zeyu21} The theoretical considerations are discussed in the next section along with introduction of the model. The practical results of these considerations are discussed in section III.
Finally, the obtained quantities are used to parameterize the Frenkel/charge transfer Holstein Hamiltonian and analyze the absorption spectra. Concluding remarks are presented in section IV.
\\~\\
\noindent\textbf{\large II. Energy \textcolor{black}{transfer} Modeling\label{sec2}}
\\~\\
\noindent\textit{The $\pi-\pi$ stacking model}
\\~\\
Figure~\ref{fig:2} demonstrates the model for studying through--space ET in layered MOFs. It is comprised of a linear array of \textit{n} SBUs with periodic boundary conditions along the non--covalent stacking direction, invoking both long--range excitonic (\textit{J}) and short--range charge transfer (\textit{T}) couplings (See the supplementary material Table S1 for specifics of the optimized SBU monomers). Both \textit{J} and \textit{T} scale with reorganization energy ($\lambda$) and are affected by interlayer spacing (\textit{S}) and displacement (\textit{D}).
Charge transport between nearest--neighbor SBUs potentially involves exciton transfer according to
\begin{equation}
  D^{*}_{n} + A_{n+1} \rightarrow D_{n} + A^{*}_{n+1}
\end{equation}\label{eq:1}
where $D^{*}_{n}$ is the $n^{th}$ excited donor SBU and $A_{n+1}$ is the nearest--neighbor acceptor SBU. According to the Bloch's theorem, the periodic function corresponds to the momentum state of a collective excitation and it is responsible for the pseudo particle behaviour.
Hence, understanding the hybrid state of layered MOFs requires an accurate evaluation of the excitonic coupling $J$ between units, in which no wave function overlap is supposed to occur\footnote{Here the assumption is that consecutive units are far enough apart that electron exchange and correlation contributions can be safely ignored.\cite{bitt09}}. In our model, whenever \textit{J} becomes sufficiently large, the reactant state $|{D^{*}_{n}A_{n+1}}\rangle$ and the product state $|{D_{n}A^{*}_{n+1}}\rangle$ may form a quantum mechanical superposition state $c_1|{D^{*}_{n}A_{n+1}}\rangle + c_2|{D_{n}A^{*}_{n+1}}\rangle$. This state is conceived as a Frenkel exciton,\cite{frenkel31}
i.e., the Hamiltonian is generalized to an arbitrary set of SBUs  along the non--covalent stacking direction of the layered MOF and expressed in a multi--particle basis set~\cite{yama11}
consisting of one and two--particle states. The excitonic eigenfunctions are then described with a wave vector $k$, which takes on the values:
\begin{equation}
k =0,\pm2\pi/n,\pm4\pi/n,...,\pi 
\end{equation}
where $k$ is in $1/S$ units. As shown in the next sections, this model allows us to differentiate the effects of long--range and short--range interactions on the spectral signatures of layered MOFs.
\textcolor{black}{
Such signatures are found to be sensitive not only to the relative phase, which by convention in this work is chosen to be consistent with periodic translational symmetry,\cite{spano12} of the intermolecular carrier transfer integrals, but also to the energy of the charge transfer state relative to the exciton state.
}
The exciton delocalization energy can be evaluated as~\cite{frenkel31}
 \begin{equation}
   E_J(k) = E_{S_1}+2Jcos(k)
 \label{ej}  
 \end{equation}
 where $E_{S_1}$ corresponds to the energy of the  $S_0 \rightarrow S_1$ electronic transition. 
 One can switch off the second term on the right hand side (RHS) of Eq.~\ref{ej} (hence  $E_J = E_{S1}$) to study the diabatic limit (i.e., no interaction between units) using this model.
 
When the electron/hole pair resides on the highest occupied molecular orbital (HOMO) and lowest unoccupied molecular orbital (LUMO) of consecutive SBUs, the charge transfer energy $E_T$ also needs to be taken into account. It can be estimated as follows~\cite{maykuhn11}
 \begin{equation}
   E_T(S) \simeq e^2/S + I_{D(A)} + A_{A(D)} + P
 \label{et}
 \end{equation}
 where the first term on the RHS of Eq.~(4) indicates the Coulomb binding energy between electron and hole separated by the distance $S$.
 Second and third terms measure, respectively, the energy to remove an electron from the donor (acceptor)
 (i.e., the ionization potential) and the electron affinity of the acceptor (donor). Finally, $P$ estimates the polarization energy of the unit. 
\textcolor{black}{
\\~\\
\noindent\textit{The long- and short-range couplings}
\\~\\
Here, we briefly introduce the techniques to evaluate couplings and reorganization energy by which the Holstein Hamiltonian is parameterized. The interested reader is referred to the supplementary material for more details. The long--range excitonic coupling \textit{J} \cite{Hennebicq:2005,Hsu:2009,You:2013,arago15}
is evaluated using atomic--centered partial charges~\cite{manc95} or atomic transition charges (ATC) $q^t$ approach which allows obtaining the transition density from \textit{ab initio} calculations. According to this approach, \textit{J} simplifies to:
  \begin{equation}
    J_{\alpha\beta} = \sum_{i,j} \frac{q^t_i q^t_j }{ \vert {\bf R}^\alpha_i - {\bf R}^\beta_j \vert}
  \end{equation}
where the indices $i,j$ run over the atomic positions of the corresponding unit,
${\bf R}^\alpha_i$ and ${\bf R}^\beta_j$ represent the spatial position of all atoms of 
units $\alpha$ and $\beta$, respectively, and $q^t_i$ and $q^t_j$ are the related partial charges:\cite{spano13}
    \begin{equation}
      q^t_i = \sqrt{2}  \sum^{N_{AO,i} }_\zeta \sum^{N_{AO}}_\eta \sum^{unocc}_q \sum^{occ}_{p} A_{pq} c^{i\zeta}_p  c^{\eta}_q S_{\zeta\eta}.
    \end{equation}
Here, the Greek indices run over the atomic orbital (AO) basis functions whereas the Latin indices run over the occupied and unoccupied molecular orbitals (MOs). The $A_{pq}$ is the coefficient obtained from configuration interaction singles (CIS) calculation and $S_{\zeta\eta}$ corresponds to the overlap matrix of atomic orbitals at the AO level. 
\\~\\
Evaluation of short--range charge transfer coupling $T$ relies on transfer integrals $t$ which express the ease of charge transfer between two SBUs, according to our model. It is well known that these integrals can be estimated to a very good approximation for 
hole (electron) transport as half the splitting of HOMO (LUMO).\cite{cornil01,kashi18} Assuming that the dimer HOMO and HOMO-1 result from the interaction of only HOMO monomers,\cite{newman04,senthi05}
the dimer Hamiltonian is calculated in the atomic orbital basis from the Roothaan Equation:
\begin{equation}
{\bf{H}}_{AO} = {\bf{SC}}E{\bf{C}}^{-1}
\end{equation}
\noindent
where matrix ${\bf{C}}$ corresponds to the supramolecular orbitals of the dimer.
Site energies and transfer integrals can be obtained respectively as the diagonal and off--diagonal matrix elements of the system Hamiltonian ${\bf{H}}^{eff}$ after L\"owdin's transformation:\cite{lowdin50}
\begin{equation}
  {\bf H} =
\begin{pmatrix}
  e_1 & t_{12} \\
   t_{12}  & e_2 
\end{pmatrix}
\xrightarrow[\text{transformation}]{\text{L\"owdin}}
  {\bf H}^{eff} =
\begin{pmatrix}
  e_1^{eff} & t_{12}^{eff} \\
   t_{12}^{eff}  & e_2^{eff} 
\end{pmatrix}\label{eq:t12}
\end{equation}
where the off--diagonal terms depend in turn on spacing $S$ and displacement $D$ between two consecutive SBUs.
Finally, we apply Huang-Rhys theory~\cite{zhang19b} to analyze the reorganization energy contributions.
After optimizing the geometries of the ground and excited states, the normal vibrational modes are calculated by diagonalizing the mass weighted Hessian. Then, a factor $\lambda_i$~\cite{scholz09} for each normal mode $i$ is obtained by projecting the difference of the excited and ground state geometries onto the normal mode coordinates. The effective Huang-Rhys factor $\lambda_{eff}$ (in the harmonic approximation) can be written as
       \begin{equation}
       %  \lambda^2_{eff} = \sum_i \lambda^2_i~~~~~~~~~~\omega_{eff}= \frac{1}{\lambda^2_{eff}}\sum_i\lambda^2_i \omega_i
         \lambda_{eff} = \sum_i \lambda_i~~~~~~~~~~\lambda_i = \frac{m_i\omega_i}{2\hbar } R^2_i
       \end{equation}\label{eq:7}
where $\omega_i$ represents the frequency of the normal mode $i$ while $R_i$ is the projection of the displacements between the equilibrium geometries of the neutral and radical--cation or radical--anion states.
$\lambda_{eff}$ is a dimensionless parameter which if multiplied by $ \hslash \omega$ returns the reorganization energy. More importantly, since $\lambda_{eff}$ characterizes the strength of the electron--phonon coupling, it affects the optical upward and downward transitions within layered MOFs.\cite{zhang19b} For the sake of brevity, in the remainder of this manuscript we refer to $\lambda_{eff}$ simply as $\lambda$.
 }
\\~\\
\noindent\textit{Interference between $J$ and $T$}
\\~\\
As will be shown later in details, $J$ and $T$ have different spatial dependencies. 
While $J$ is a long--range Coulomb coupling, the short-range coupling  $T$ depends on the interlayer wavefunction overlap between adjacent units. This overlap is governed by the nodal patterns of the frontier molecular orbitals from which the electron ($t_e$) and hole ($t_h$) transfer parameters are obtained, by solving Eq.~\ref{eq:t12} for pairs of LUMO/LUMO or HOMO/LUMO orbitals, respectively. Hence, sliding SBUs compared to each other decides the phase of the $t_e\times t_h$ product.
The effective coupling provided by the interference of $T$ with $J$ can be either constructive or destructive, according to the expression:\cite{harcourt94}
\begin{equation}\label{eq:J_eff}
    J_{eff} = J + T
\end{equation}
where
\begin{equation}
    T = -2 \frac{t_e\times t_h}{E_{T} - E_{S_1}}\label{eq:T}
\end{equation}
Eq.~\ref{eq:T} implies a two-step superexchange process~\cite{harcourt96} in which the exciton moves from one SBU to its neighbour via a virtual charge transfer state. In our model we assume $E_T > E_{S_1}$. Hence $T$ can be positive or negative depending on the signs of $t_e\times t_h$. 
\textcolor{black}{Furthermore, whenever transfer integrals and  $J$ are positive, long and short-range couplings have opposite signs.}
Similar to observations made for the uniaxial packing of organic molecules,\cite{zang08} the relative sign between $t_e$ and $t_h$ controls the J- and H-like\footnote{Here \textit{J}  stands for induced red shift~\cite{jelley36} whereas \textit{H} stands for induced blue shift~\cite{allolio18} as a result of the aggregation.} photophysical properties in layered MOF aggregates, a point that will be further explored in the next section.
\\~\\
\noindent\textit{The Holstein Hamiltonian: Modeling the interrelation between $J$ and $T$}
\\~\\
The Holstein Hamiltonian~\cite{holst59,spano10,spano17} used in this study is comprised of three parts:
\begin{equation}\label{eq:h}
  H= H_{J} + H_{T} + H_{POL} 
\end{equation}
the Frenkel term $H_{J}$  includes the vibronic coupling:
\begin{multline}
  H_{J} = \hbar \omega_{vib} \sum_n b^{\dagger}_n b_n + \hbar \omega_{0-0} \sum_n |c_n,a_n\rangle\langle c_{\textcolor{black}{n}},a_{\textcolor{black}{n}}|\\ 
  +\hbar \omega_{vib} \lambda\sum_{n} \left( b^{\dagger}_n + b_n + \lambda \right) |c_n, a_{n}\rangle \langle c_n,a_{n}|\\
  +\sum_{n,m} J_{n,m} |c_n,a_n\rangle\langle c_m,a_m|
\end{multline}
\textcolor{black}{
where the first sum corresponds to pure vibrational energy contribution.
The ket in the second sum denotes that monomer $n$ hosts a Frenkel exciton while all other molecules remain in the electronic ground state.}
The third term corresponds to contribution from the local vibronic coupling which is related to the Huang--Rhys factor via $\lambda^{2}$.
%(measured by the dimensionless quantity $\lambda$).
\textcolor{black}{
The last term monitors the long–range Coulomb couplings between two monomers $n$ and $m$ as the donor and acceptor, respectively.}
{\textcolor{black}{The}} frequency $\omega_{0-0}$ corresponds to the 0-0 transition  and $b^{\dagger}_{n}(b_{n})$ creates (annihilates) a vibrational excitation with energy $\omega_{vib}$ on the $n^{th}$ monomer.\\
The second term of Eq.~\ref{eq:h} accounts for electron and hole transfers:
\begin{multline}
  H_{T} =   t_e\sum_{n,s} ( |c_n, a_{n+s}\rangle\langle c_n,a_{n+s+1}| \\
  + |c_n, a_{n+s}\rangle\langle c_n,a_{n+s-1}| ) 
  + t_h \sum_{n,s} ( |c_n, a_{n+s}\rangle\langle c_{n+1},a_{n+s}| \\+ |c_n, a_{n+s}\rangle\langle c_{n-1},a_{n+s}|)
\end{multline}
where $t_e$ and $t_h$ represent the transfer integrals between nearest neighbour SBUs in our model, see Figure~\ref{fig:2}.
\\
The last term in Eq.~\ref{eq:h} includes the energies of the polaronic states and their associated vibronic couplings:
\begin{multline}\label{eq:pol}
  H_{POL} = \sum_{n,s \neq 0} \hbar\omega_{T}|c_n,a_{n+s}\rangle\langle c_n,a_{n+s}|\\
  +\hbar \omega_{vib} \lambda_{+}\sum_{n,s \neq 0} \left( b^{\dagger}_n + b_n + \lambda_{+} \right) |c_n, a_{n+s}\rangle\langle c_n,a_{n+s}|\\
  + \hbar \omega_{vib} \lambda_{-}\sum_{n,s \neq 0} \left( b^{\dagger}_{n+s} + b_{n+s} + \lambda_{-} \right) |c_n, a_{n+s}\rangle\langle c_n,a_{n+s}|
\end{multline}
where the $\omega_{T}$ energy is a function of the separation ($S$) between the 
hole on monomer $n$ and the electron
on monomer $n + s$, whereas the second and the third term of Eq.~\ref{eq:pol} monitor the local vibrational coupling
quantified by the dimensionless $\lambda_{(+/-)}$ factors. A comprehensive study of the effect of structural changes of SBU and OLU on these factors is provided in the supplementary material.
\\~\\
\noindent\textit{Spectral analysis}
\\~\\
\textcolor{black}{
The Hamiltonian in Eq.~\ref{eq:h} is diagonalized numerically, yielding  eigenstates and energies from which the absorption spectra are 
calculated using the equation
\begin{equation}
  A(E) = \sum_i {\bf f_i} \Gamma_i(E).
\end{equation}
Here, $\Gamma_i$ is a line broadening function and $\bf{f_i}$ corresponds to the oscillator strength of the $i^{th}$ excited state which is given by
the product of the transition energy and the square of the transition dipole moment
\begin{equation}
  {\bf f_i} = \frac{1}{\mu^2} \left|\braket{G\left|\hat{{\mu}}\right| \Psi_i} \right|^2\label{eq:fi}.
\end{equation}
In Eq.~\ref{eq:fi}, $\ket{\Psi_i}$ is the $i^{th}$ eigenvector of the Hamiltonian in
Eq.~\ref{eq:h} with energy $E_i$, $\ket{G}$ represents the vibrationless ground state and $\hat{\mu}$
is the transition dipole moment operator
\begin{equation}
  \hat{\mu} = \sum_n {\bf \mu_n} \ket{g_n}\bra{e_n} + H.c.
\end{equation}
The line broadening function is taken to be a normalized Gaussian 
\begin{equation}
  \Gamma_i(E) = \frac{1}{\sqrt{2\sigma^2_i \pi}}exp\left[- \frac{\left(E - E_i \right)^2}{2 \sigma^2_i}  \right]
\end{equation}
where $\sigma_i$ is the line width of the $i^{th}$ transition. We have used this procedure and the Hamiltonian in Eq.~\ref{eq:h} to reproduce the experimental absorption spectrum of Ni$_3$(HITP)$_2$ reported in the supplementary material of Ref. \onlinecite{shebe14}. To make our stacking model compatible with Ref. \onlinecite{shebe14}, we set $S=3.3$ \AA~ and $D=1.85$ \AA. As can be seen in Figure \ref{fig:ex-th}, a very good agreement is obtained from our simulations and the experimental absorption 
\begin{figure}[h]
  \centering
    \includegraphics[width=0.99\linewidth]{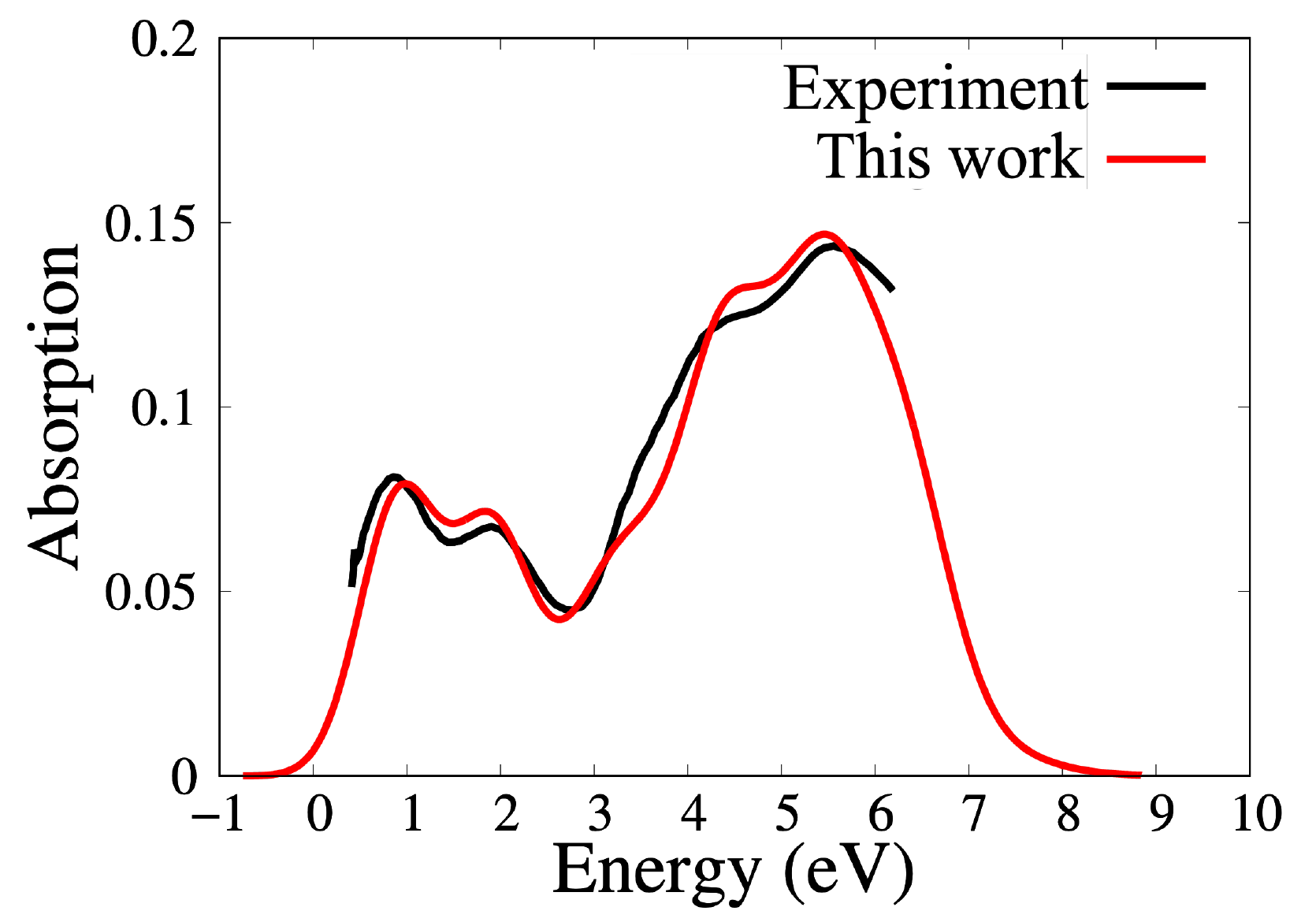}
    \caption{ Comparison between the experimental adsorption spectrum\cite{shebe14} of Ni$_3$(HITP)$_2$ with the one obtained from our model Hamiltonian. The set of parameters used in these simulations are $J=1100$, $t_h=910$, $t_e=875$, $E_{S_1}=2500$, and $E_T=3900$ $cm^{-1}$. The absorption line width has been increased to a value of 650 $cm^{-1}$.
    }\label{fig:ex-th}
\end{figure}
\begin{figure*}[t]
  \centering
    \includegraphics[width=0.99\linewidth]{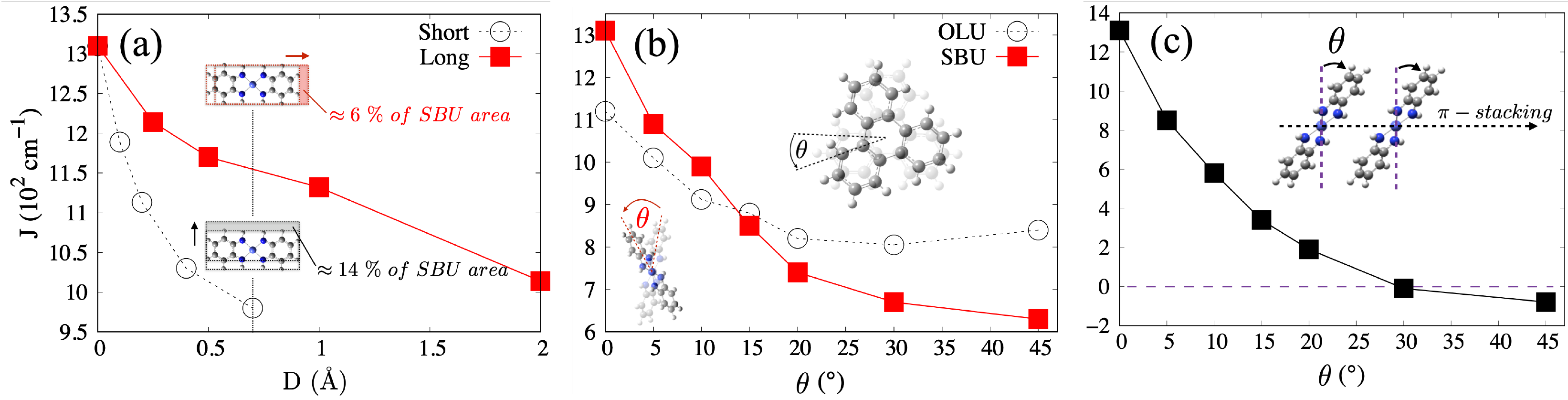}
    \caption{
      $J$ profiles obtained from the interaction between the two building blocks of Ni$_3$(HITP)$_2$ layers. The couplings are shown (a) as a function of the displacement in \textit{ab} plane along the long and short molecular axes, (b) as a function of the rotational angle between the two monomers, representing the twisting of adjacent layers, and (c) as a function of the tilt angle between two SBUs along the $\pi-\pi$ stacking direction. The center of mass between two adjacent monomers is fixed at a distance of $3.3$~\AA.
      }\label{fig3_new}
\end{figure*}
spectrum of Ni$_3$(HITP)$_2$ from Ref. \onlinecite{shebe14} in reproducing the position and shape of the peaks. In the next sections we will use our methodology to provide insights on how different possible motions of the layers  affect the photophysical properties of Ni$_3$(HITP)$_2$. These insights will be useful in proposing ways to engineer layered MOFs with desirable properties.} 
\\~\\
\noindent\textbf{\large III. Results and Discussion\label{R&D}}
\\~\\
\noindent\textit{How $J$ is affected by in--plane displacement, twisting, and rotations}
\\~\\
It is known that the stacking of aromatic building blocks can turn Ni$_3$(HITP)$_2$ into 
a semiconducting material with peculiar optoelectronic properties.\cite{foster18} The reason behind it, as we have shown recently,\cite{zeyu21} is the constant motions of the layers which is not only related to the change of interlayer distance in the stacking direction but also in-plane displacement as well as twisting of the layers. \textcolor{black}{We have shown\cite{zeyu21} that PBE--D3 optimized Ni$_3$(HITP)$_2$ layers at 0 K are displaced with respect to each other in the \textit{ab} plane which is consistent with the previously characterized slipped-parallel stacking configuration for this material.\cite{shebe14,chen:2015} Our subsequent \textit{ab initio} molecular dynamics (AIMD) simulations\cite{zeyu21} at 293 K revealed that in-plane displacement is a continuous movement.} Recently, Tao~$et~al.$~\cite{tao19} showed that the structure of layered MOFs obtained by randomly stacking is the one where adjacent layers are twisted compared to each other, while untwisted mode is generated by a procedure which consists of heating the twisted nanosheets. Although it is possible to control the stacking pattern and engineer optoelectronic properties by strong interlayer interactions and functional groups,\cite{kuc20} to the best of our knowledge the twisting and displacement control/engineering are still in their infancy. Chemical modification may permit to tune displacements of the layers compared to each other, for instance by substitution of the terminal groups~\cite{hart14} or through lattice strain.\cite{giri11}
The central question here is to what degree the mutual twisting and displacement between different building blocks in layered MOFs may impact the ET along the $\pi-\pi$ stacking direction. \textcolor{black}{In Figure~\ref{fig3_new}a the ATC procedure, outlined in previous section, has been applied to calculate the excitonic couplings $J$ as a function of the displacement (D) of the two monomers compared to each other in two different directions, i.e. along the short and long molecular axes as shown in the insets. As can be seen, displacement along the short molecular axis leads to a faster decay of $J$ coupling as a result of more significant loss of overlap between the two monomers. The insets in Figure~\ref{fig3_new}a demonstrate that at the same D, the two monomers loose a bigger overlap area along the short molecular axis than the long axis.} 

In Figure~\ref{fig3_new}b, the ATC procedure has been applied to calculate the excitonic couplings $J$ as a function of the in--plane rotational angle between two adjacent SBUs or OLUs. (see supplementary material, Table S2, for a list of the transition charge dipole values for triphenylene  systems  as  a  function  of  the  rotation  angle  around  the  OLU  center.)
In particular, one unit of the dimer is kept fixed whereas the other unit undergoes a rotation around its center, starting  from the native configuration  (that is $0$\degree) until $45$\degree. 
In native configuration, the SBU shows a larger $J$ value compared to OLU. As the rotational angle increases, SBU and OLU demonstrate different angle dependency of their $J$ values. The $J$ profile of SBU spans $\sim 700~cm^{-1}$ while that of OLU spans only $\sim 300~cm^{-1}$. The excitonic coupling profile of OLU shows an almost plateau feature after rotational angle of $20$\degree~with a barely noticeable minimum at $30$\degree. On the other hand, excitonic coupling between the two SBUs constantly decreases by increasing the rotational angle. Overall, irrespective of the building block, we can notice a broad region around the native configuration where the coupling has a positive value. More importantly, the slope difference between the two profiles may provide crucial information in designing frameworks with specific $J$ characteristics. 

\begin{figure}[t]
  \centering
    \includegraphics[width=0.99\linewidth]{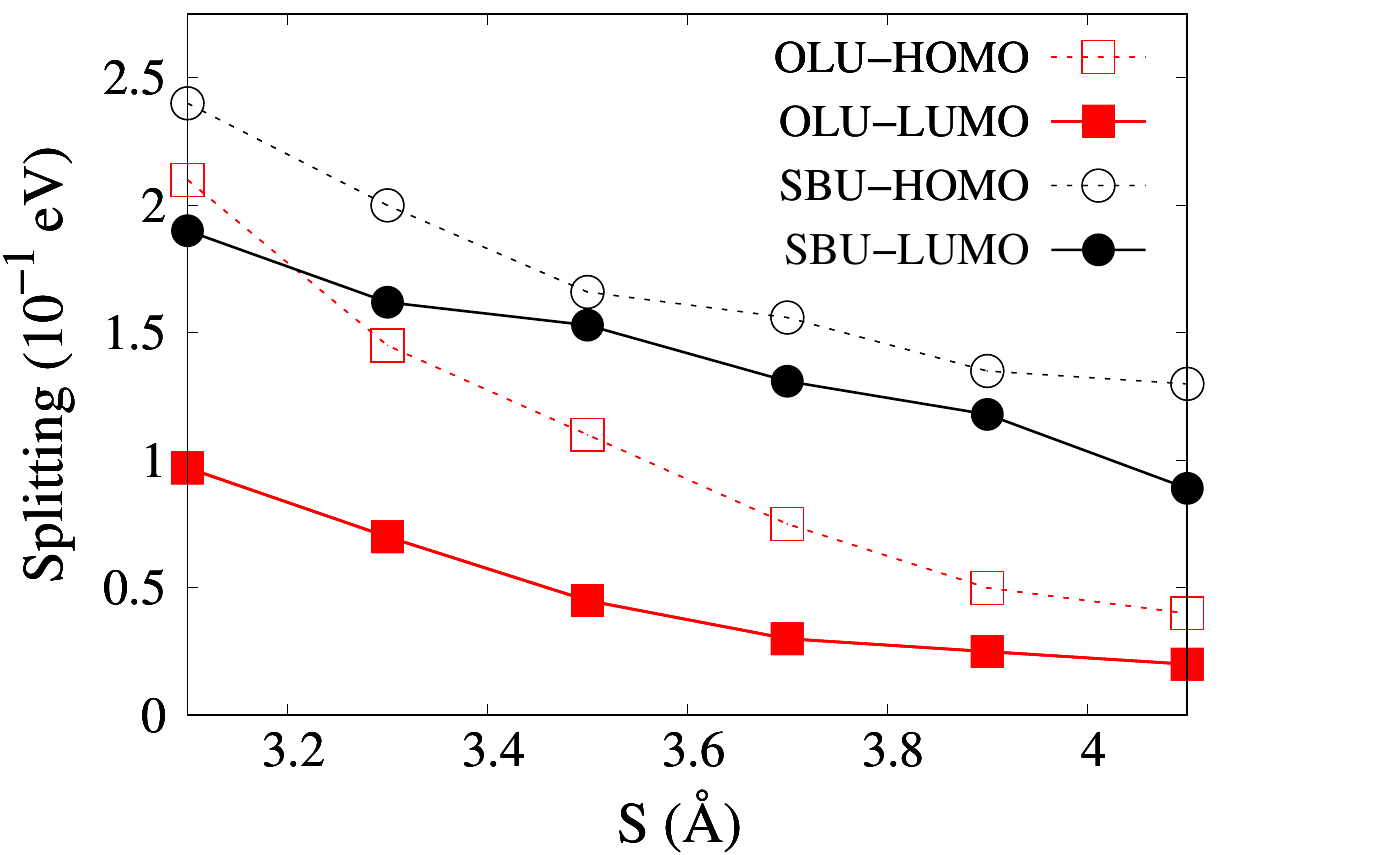}
    \caption{
      HOMO--LUMO splitting ($t$) for SBU (black) and OLU (red) increasing the spacing $S$ along 
      the stacking direction as represented  in Figure~\ref{fig:2}.
  }\label{fig:6}
\end{figure}

Our recent AIMD simulations\cite{zeyu21} at 293 K showed that the equilibrated layers of Ni$_3$(HITP)$_2$ adopt a stepped geometry where the SBUs are tilted with respect to the \textit{ab} plane. This is opposed to fully planar layers that are normally obtained from \textit{static} electronic structure calculations at 0 K. While engineering optoelectronic properties based on tilt angles in the stacking direction is still beyond reach, it is interesting to see the effect of this geometrical aspect of Ni$_3$(HITP)$_2$ layers on the excitonic coupling. Figure \ref{fig3_new}c demonstrates the $J$ profile of two adjacent SBUs as a function of increasing the tilt angle along the stacking direction. Change of the tilt angle, which occurs at room temperature vs. 0 K, results in a drastic change of excitonic coupling. In contrast to the layer displacement case, here we notice a change in the sign of $J$. Indeed, at about $45$\degree~the dimer switches from a head--to--head to a head--to--tail configuration, according to Kasha's theory.\cite{guerrini19}
%\begin{figure}[ht]
%  \centering
%    \includegraphics[width=0.99\linewidth]{fig5}
%    \caption{$J$ profile increasing the tilt angle between two SBUs along the $\pi-\pi$ stacking direction and \textcolor{red}{separated by a distance of $3.3$~\AA}.
%    }
%    \label{fig:5}
%\end{figure}
\\~\\
\noindent\textit{Energetic splittings of the frontier molecular orbitals}
\\~\\
It has been shown that the conductivity in Ni$_3$(HITP)$_2$ decreases with $S$ and increases with $D$.\cite{chen20}
 The latter is rather counter intuitive if we consider that the interlayer non--covalent interaction is based on $\pi-\pi$ stacking. Insights into the ET behaviour of Ni$_3$(HITP)$_2$ can be obtained by analyzing the HOMO and LUMO energetic splittings as an estimate of the magnitude of transfer integrals in the building block dimers introduced in the previous section.
 Previous studies on triphenylene derivatives have shown that eclipsed configuration provides the largest electronic
 interactions between adjacent molecules/oligomers and highest value of transfer integrals.\cite{senthi03}
 The splitting depends on the initial shapes of HOMO and LUMO wavefunctions,\cite{kashi18} i.e.
 the number of nodes which determines the initial bonding/antibonding contributions. Hence, here we examine the energetic splitting profiles of SBU and OLU dimers as a function of $S$ starting from an eclipsed configuration as shown in Figure~\ref{fig:6}. 
 SBU splittings decrease less rapidly than OLU because of the spatial extent (and the resulting wavefunction overlap) 
of the $d_z$ orbitals of Ni atoms in comparison to $p_z$ orbitals of C atoms in OLU. Since the wavefunction decays exponentially with $S$, these profiles show that the transfer integral between two OLUs should already be negligible at $S$ values larger than $~4.1$~\AA. Remembering that the experimentally and theoretically evaluated interlayer distance of Ni$_3$(HITP)$_2$ is $\sim$3.3~\AA, one can conclude that there is no charge transfer coupling between OLUs beyond the nearest-neighbour layer. While the splitting, and the resulted $T$ magnitude, is higher for SBU, one should remember that  the intermolecular overlap of the electronic wavefunctions also depends delicately on the displacement of SBUs compared to each other. Sliding and twisting of layers in Ni$_3$(HITP)$_2$, which is established through both experimental and theoretical studies, will drastically affect the energy splitting. We will further develop this concept by studying the photophysical properties of SBUs in the next section.  
 \begin{figure*}[ht]
  \centering
    \includegraphics[width=0.99\linewidth]{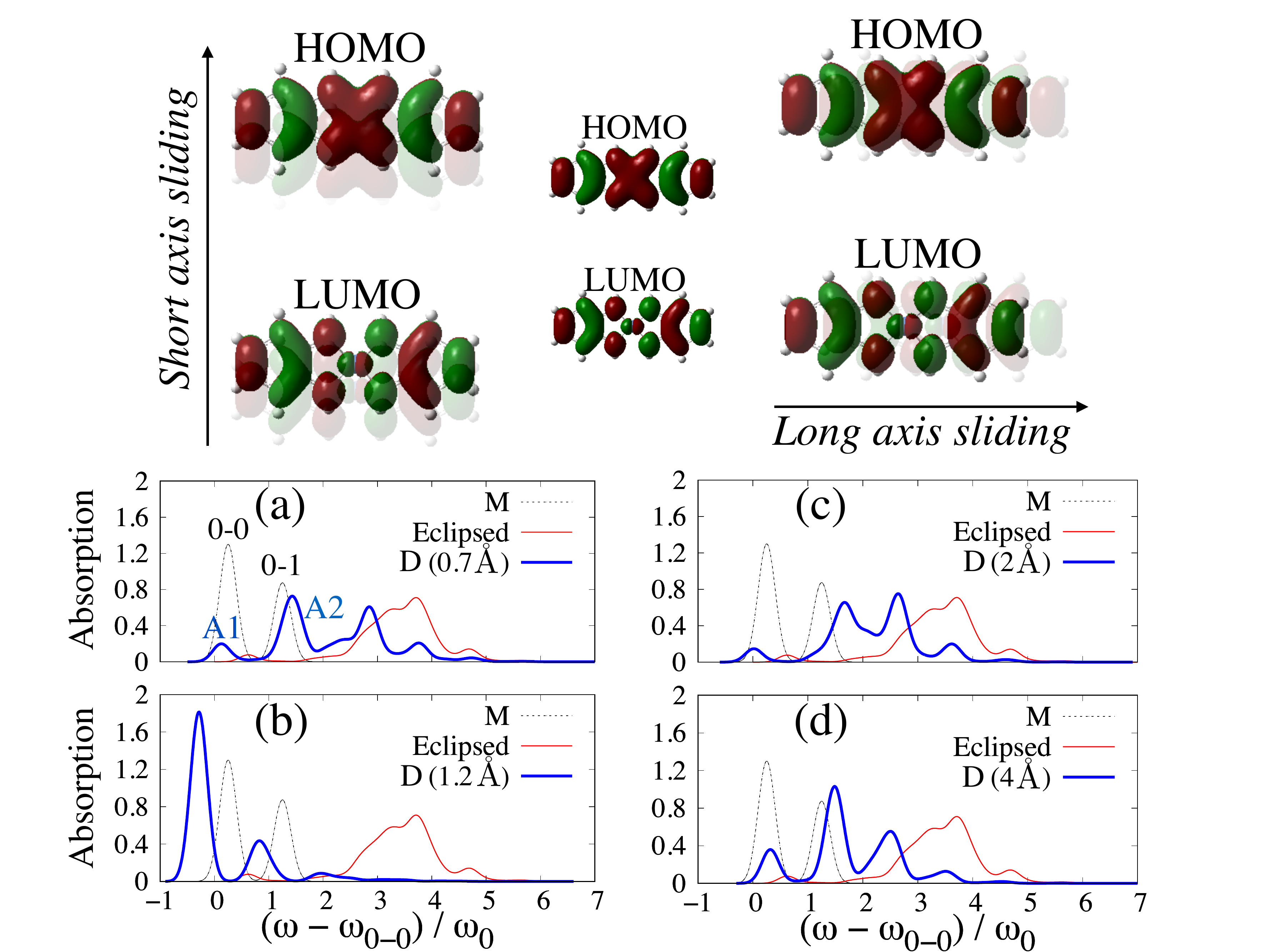}
  \caption{ Short (a-b) and long (c-d) axis sliding effect on the photophysical properties of SBU.
  In each panel, the spectra for the monomer M (black, dashed) and fully eclipsed (red) are reported.
  }\label{fig:7}
\end{figure*}
 \\~\\
\noindent\textit{The vibronic signature: layered MOF material screening for certain photoelectronic characteristics}
\\~\\
As introduced in Section II, in order to correlate a nanoscopic $J/\lambda$ or  $T/\lambda$ event to mesoscopic transport, exciton diffusion is modeled as an ensemble of self-carrier/energy transfer hopping events on a stack arrangement of SBUs accompanied by nuclear relaxation and rearrangement. The stack is composed of $n=30$ SBU monomers since it is the value of $n$ by which the absorption spectra converge. As suggested by Eqs.~\ref{eq:J_eff}-\ref{eq:T}, the spectral response relies on the competition between $J$ vs. $T$, the latter related to phase and magnitude of the transfer integrals. Figure~\ref{fig:7} shows the absorption spectra of the stack of SBUs subject to sliding along the short molecular axis in plots (a) and (b) as well as sliding along the long molecular axis in plots (c) and (d). The full set of excitonic and charge transfer coupling parameters, obtained from electronic structure calculations as discussed in the previous sections, is reported in Table 1. For comparison, the spectra of the monomer and the completely eclipsed dimer are also displayed in each panel in Figure~\ref{fig:7} . 
\begin{table}
   \begin{center}
     \begin{tabular}{cccc}
 \hline
               ~~~&~~~{\bf $J$}~~~&~~~{\bf $t_h$}~~~&~~~{\bf $t_e$} \\
 \hline
       Eclipsed~~~&~~~1310~~~&~~~784~~~&~~~-772 \\
               ~~~&~~~{\bf Short axis sliding}~~~&~~~&     \\
       0.7 \AA~~~&~~~980~~~&~~~512~~~&~~~293  \\
       1.2 \AA~~~&~~~-605~~~&~~~192~~~&~~~20   \\
               ~~~&~~~{\bf Long axis sliding}~~~&~~~&     \\
        2 \AA~~~&~~~1014~~~&~~~-689~~~&~~~290  \\
        4 \AA~~~&~~~714~~~&~~~-287~~~&~~~-105   \\
 \hline
 \textcolor{black}{~~~$E_{S_1}$~=~600}~~~&~~~\textcolor{black}{$E_{T}$~=~6000}~~~&~~~\textcolor{black}{$\sigma$~=~250}\\
  \hline
  \end{tabular}
     \caption{
       Full set of parameter values (in wavenumbers) used in Figure ~\ref{fig:7}. \textcolor{black}{The energies of diabatic CT ($E_{S_1}$) and Frenkel states ($E_T$) as well as the absorption line width ($\sigma$) are also provided.}}\label{tab:4} 
   \end{center}
 \end{table}
The intensity difference between the first two vibronic peaks (0-0 and 0-1) in the monomer is due to a $\lambda$ value of $0.82$ (i.e. they would have similar height if $\lambda$ were set to unity). \textcolor{black}{The} red profile corresponds to the eclipsed stack for which $\lambda_+$ and $\lambda_-$ values are $0.46$ and $0.43$, respectively. Energies are scaled with respect to the elongation of a symmetric stretching mode (subsequent to $S_0 \rightarrow S_1$ optical excitation) of SBU with energy 1568 $cm^{-1}$ and reported in vibrational quanta units. 
We monitor not only the shift in the position of the spectral lines but also how the ratio of the first two vibronic peaks (A1/A2) deviates from that of the monomer.
\textcolor{black}{
As it has been already reported,\cite{spano05,spano10} the ratio of the first two vibronic
peaks (A$_1$ and A$_2$) of the absorption spectrum depends on the nature of the Coulombic coupling.
In particular, this ratio increases with the magnitude of the intermolecular coupling in J-aggregates
 ($J_{Coul} < 0$) and decreases in H-aggregates ($J_{Coul} > 0$).
}
As we are concerned with the limit in which transfer integrals and vibrational energies are small compared to the energetic separation between $T$ and $J$ states, the $E_T/E_{S_{1}}$ ratio (see Eqs.~\ref{ej} and~\ref{et}) is initially settled to a perturbative regime value of one order of magnitude.
 \begin{figure}[ht]
   \centering
     \includegraphics[width=0.99\linewidth]{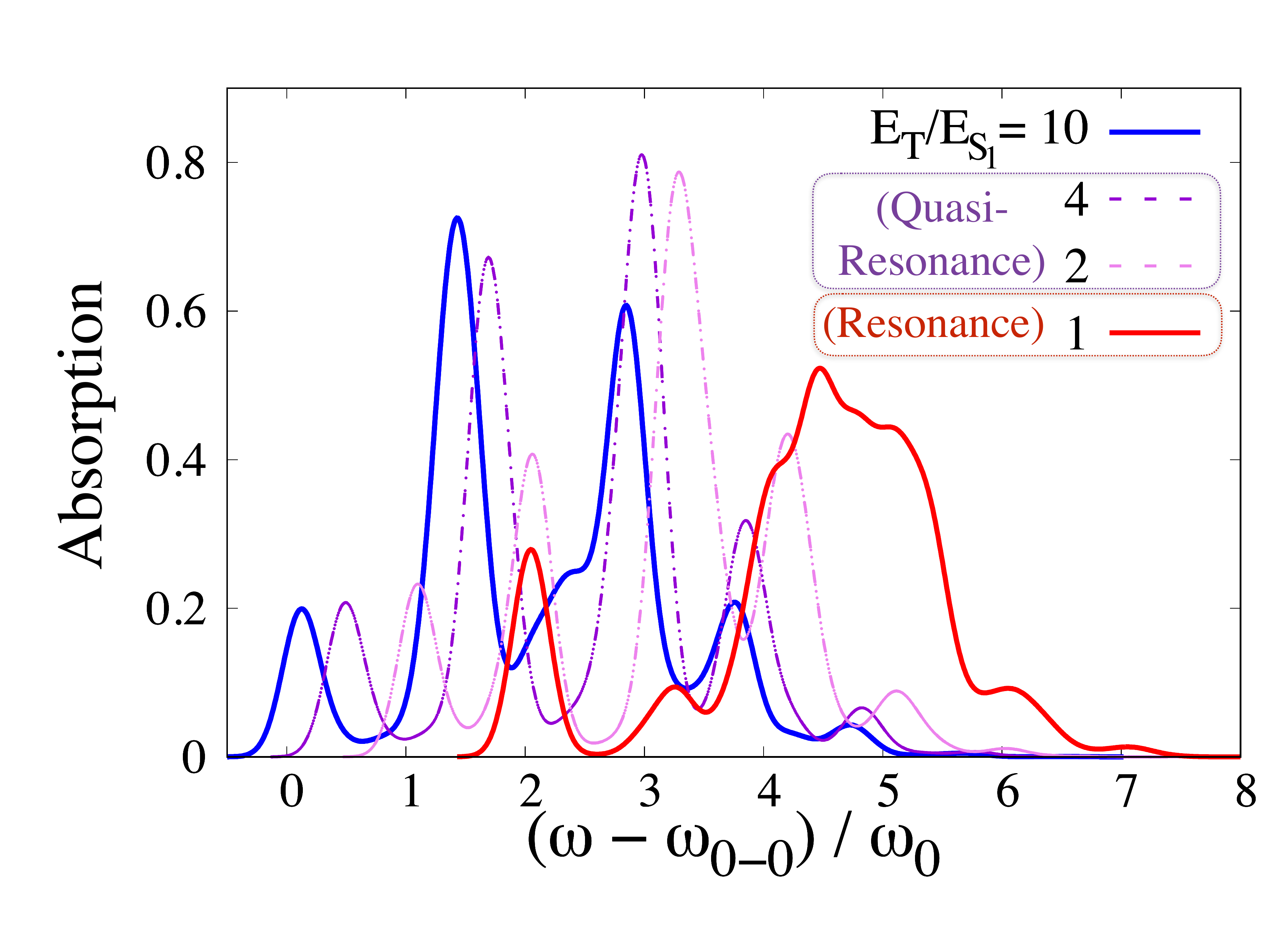}
   \caption{Comparison between the absorption spectra obtained with a large
   $E_T/E_{S_{1}}$ ratio (solid blue line) as well as at the resonance value (solid red line).
   Dashed violet profiles correspond to intermediate $E_T/E_{S_{1}}$ values.
   For clarity, spectra are shifted of one vibration quanta decreasing the $E_T/E_{S_{1}}$ ratio.
   }\label{fig:8}
 \end{figure}
 
 \begin{figure*}[!htb]
   \centering
     \includegraphics[width=0.99\linewidth]{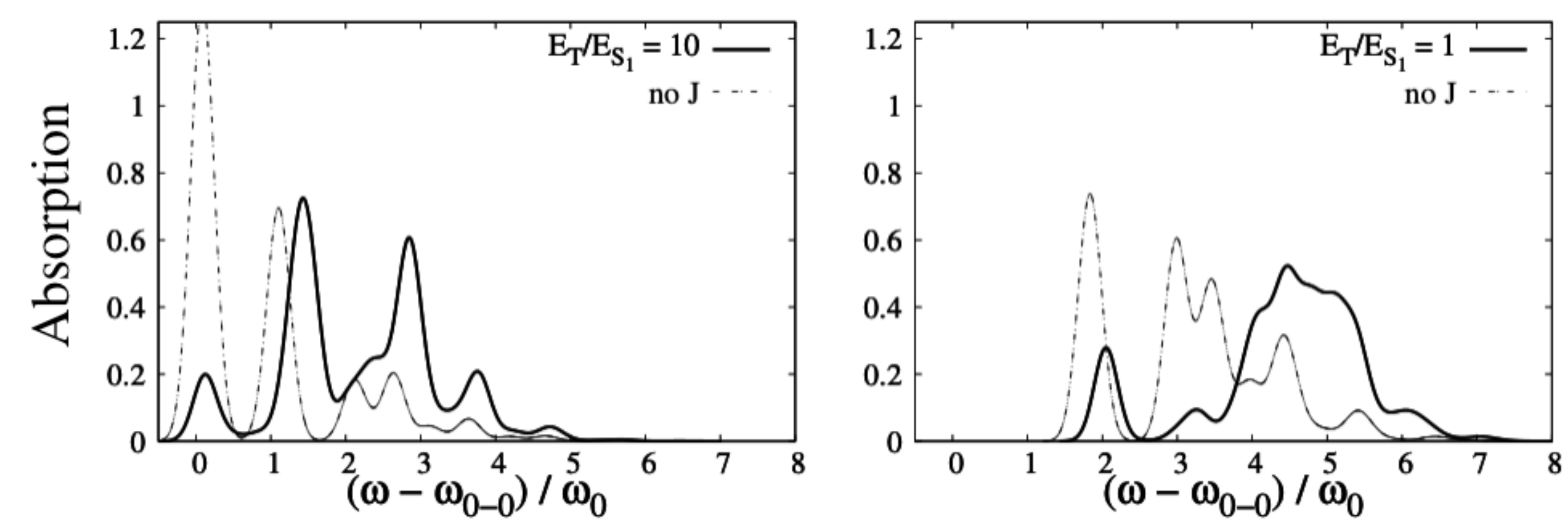}
     \caption{Spectral study by switching off the Coulomb interaction $J$ in Eq. (3). Solid profiles are taken from Figure 9 whereas the dashed lines are the ones obtained after switching off $J$.}\label{fig:9}
 \end{figure*}
 Sliding as big as 0.7 \AA~along the short axis (Figure~\ref{fig:7}(a)) causes a \textcolor{black}{red} shift in the absorption spectrum compared to the eclipsed units. The Coulomb coupling is H-like and the A1/A2 ratio drops below the 0-0/0-1 ratio of the monomer. $J$ and $T$ have a destructive phase in this case and the former, although it decreases by more than 25\% compared to the eclipsed case, dominates over the latter.
 The nodal patterns of the frontier molecular orbitals (see the top of Figure~\ref{fig:7} for a pictorial representation) impact the magnitude of the transfer integrals. In particular, $t_e$ decreases more noticeably than $t_h$ due to the LUMO contribution of the metal center. When the two units slide as big as 1.2 \AA~, the sign of $J$ becomes negative, Table 1, leading to a constructive interference between $J$ and $T$, Figure~\ref{fig:7}(b). This leads to a reinforced coupling and a pronounced J-like behaviour with A1/A2 becoming greater than that of a monomer. Hence, a slide along the short axis can lead to either H-like or J-like behaviour depending on the sign of $J$.

Now, let us examine the photophysical behaviour when sliding occurs along the long molecular axis. Due to the specific form of HOMO and LUMO orbitals (see the pictorial representation on top of Figure~\ref{fig:7}), every time that the displacement exceeds a bond distance, a nodal plane forms between the two SBUs. After sliding as big as $2$ \AA, Figure~\ref{fig:7}(c), $t_h$ becomes negative but not $t_e$. $J$ and $T$ have constructive phase here, and the larger H-like long-range coupling dominates the photophysical response.
This generates a \textcolor{black}{red shift with respect to the eclipsed configuration,} similar to Figure~\ref{fig:7}(a) although a difference in the profiles is visible at $\sim2$ vibration quanta. Also, the \textcolor{black}{red} shift in panel (c) shows a larger spacing  between A1 and A2 peaks if compared to the \textcolor{black}{red} shift of panel (a). Finally, Figure~\ref{fig:7}(d) shows the spectral response corresponding to the sliding limit along the long molecular axis during which $J$ retains its sign, i.e. within $4$~\AA.\textcolor{black}{~Both $t_e$ and $t_h$ have negative signs and the overall spectrum shows a blue shift signature with respect to the monomer.} 

Next, we move to the case where the energetic difference between $T$ and $J$ states decreases. In Figure~\ref{fig:8} for a sliding as big as $0.7$~\AA~we show the spectral consequence moving from a perturbative regime (i.e., $E_T/E_{S_{1}}\approx10$) toward a resonance regime ($E_T/E_{S_{1}}\approx1$). Departure from perturbative limit leads to an overall blue shift in absorption spectra. The A1/A2 ratio displays H-like characteristics only in quasi-resonance regime. At the energetic resonance this ratio becomes greater than one which is a J-like character. Further, novel spectral signatures arise, importantly a broad peak centered approximately at 5 vibrational quanta.
This clearly indicates that the photophysical response over ET in layered MOFs may be dramatically affected by not only the spatial dependence of the SBUs and their relative orientation but also by the energetic separation between charge transfer and Coulomb coupling.

 Finally, in order to focus uniquely on the impact of charge transfer interactions, we examined the spectral lines in the diabatic limit, i.e. by switching off the second term on the RHS of Eq.~\ref{ej}, i.e., no Coulomb interactions between SBUs. The two perturbative and resonance limits, i.e., $E_T/E_{S_{1}}=10$ and 1, are chosen to be compared in the case of a 0.7 \AA~slide along the short molecular axis. These two cases correspond to the blue and red profiles in Figure~\ref{fig:8}. The original spectrum (solid line) and the one after switching off $J$ interactions (dashed line) are shown in Figure \ref{fig:9} for both cases. As can be inferred from the A1/A2 ratio exceeding unity, dashed profiles show J-like features in both cases. Since the 0.7 \AA~slide nearly resembles eclipsed form, according to the understanding provided by Kasha's model we would expect a decrease of the ratio of the two first vibronic peaks. Instead, this result illustrates that the system based on the interaction between SBUs may display photophysical behaviour opposite to what is expected based on the geometry.
 For engineering purposes, one may exploit the spatial  dependencies of $J$ and $T$ to design layered MOFs for specific optoelectronic applications in which (for instance) the Coulomb coupling is H-like, but the photophysics are J-like due to a dominant J-like $T$ interaction between building units.
\begin{figure*}[ht]
  \centering
    \includegraphics[width=0.99\linewidth]{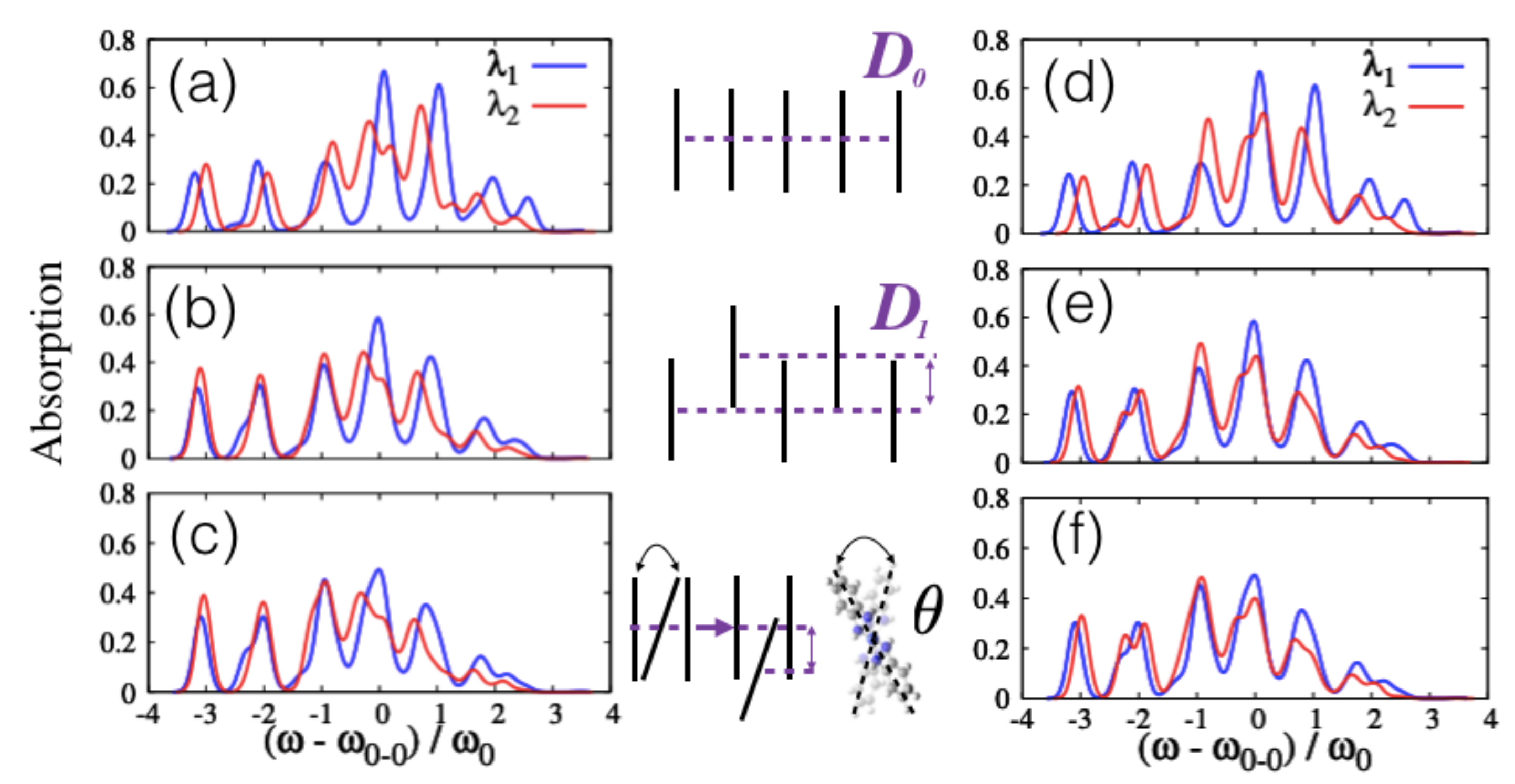}
\caption{Absorption spectra obtained from the parametrization of the Hamiltonian, Eq.~\ref{eq:h}, increasing the electron-phonon coupling (i.e., $\lambda_1 \rightarrow \lambda_2$). (a)(d) eclipsed non-covalent aggregate in the stacking direction; (b)(e) sliding between consecutive units; (c)(f) sliding $plus$ rotation
between consecutive units. As in Figure~\ref{fig:7}, spectra are scaled with respect to the spectral frequency of the centroid $\omega_0$.}\label{fig:10}
\end{figure*}
\\~\\
\noindent\textit{Modeling the electron-phonon coupling by monitoring the vibronic progression}
\\~\\
In the previous section we examined how the spectral response is related to the short and long-range interactions between SBUs. One should not forget that this response scales with the reorganization energy involved during an ET event, a factor which may potentially impact the transfer, especially when consecutive building units are displaced from an eclipsed arrangement. As introduced in section II, Huang–Rhys factor $\lambda$ takes into account electron–phonon coupling strength due to lattice relaxation, where the latter refers to the change in equilibrium atomic positions between the initial and final state involved in the optical transition.
The design or at least understanding the structure-function relationship of a new layered scaffold may start by monitoring the spectral vibronic progression by applying slight variations in the electron-phonon coupling during the sliding motion. In such progression, the energies of the normal vibrational modes that take part in the nuclear reorganization determine the vibronic peak spacing.
Starting from the eclipsed model, Figure ~\ref{fig:10}(a-c) shows the changes in intensity and spacing of the vibronic peaks due to sliding and rotation subject to different values of $\lambda$.
Specifically, in these simulations we set $\lambda$ near unity, i.e. it varies from $\lambda_1 = 0.82$ to $\lambda_2 = 0.86$. 
$S$ is kept fixed while $D$ is varied from completely eclipsed (i.e., $D_0 = 0$ \AA~in Figure \ref{fig:10}(a) and \ref{fig:10}(d)) to a slide of $D_1= 1$ \AA~in Figure \ref{fig:10}(b) and \ref{fig:10}(e) while a rotation as big as $\theta=20^{\circ}$ between consecutive SBU units relative to one another,  about the axis connecting their mass centers, is added to the sliding in Figure \ref{fig:10}(c) and \ref{fig:10}(f). 
The eclipsed formation in Figure \ref{fig:10}(a) shows that the spectral centroid is slightly red-shifted by increasing $\lambda$. Together with a broader spectrum, the red profile displays the emergence of a split
of the central peak, leading to two peaks symmetrically positioned around zero. These are also the common observation in Figure \ref{fig:10}b, except that now the ratio of the first two vibronic peaks in higher value of $\lambda$ is associated with a J-like aggregate form. This demonstrates how a small change in the electron-phonon coupling may be employed to engineer a dramatic variation in the spectral response. When the rotation is added to the sliding the interference scenery between $J$ and $T$ as a function of rotation angle is expected to be dependent on the nodal structures of the frontier orbitals. Figure \ref{fig:10}(c) shows that the central vibronic peaks increase their spacing irregularity. Further, one may appreciate a modest redistribution of the relative areas of the peaks, which may suggest a reduced spread of the wavefunction.\cite{mayers18} It can be suggested that for design purposes the effect of rotation on coupling sources may be fine-tuned by adjusting the denominator of Eq.~\ref{eq:T}.

Finally, in Figure ~\ref{fig:10}(d,e,f) a random field to mimic a certain degree of thermally-induced disorder has been introduced. As previously proposed for perovskites,\cite{motta16} this was achieved by adding a small random fluctuation, uniformly distributed in an interval $\left[0, \epsilon \right]$, to the onsite energy of every unit of the stacked model. Introduction of this fluctuation slightly increases the match between blue and red profiles in Figure \ref{fig:10}(e) and \ref{fig:10}(f) in comparison to Figure \ref{fig:10}(b) and \ref{fig:10}(c). 
Optimizing ET in $operando$ conditions is crucial in the design of efficient layered MOF based devices, which requires that the absorbed energy (i.e. the exciton) be carried to regions within the device that may enhance charge separation before it is lost via either radiative or non-radiative decay processes. On one hand, thermally-induced vibrations introduce an additional loss of coherence~\cite{yutao20} among the interacting units along the $\pi-\pi$ stacking which leads to the reduction of effective bandwidths (i.e., the transfer integrals). On the other hand, as previously shown on hybrid perovskites,\cite{even14} it is possible that the molecular motions create defects and hence increase the exciton lifetime by localizing electrons and holes, i.e., reducing their recombination rate.\cite{ma15} Both of these effects should be taken into account in quantitative evaluation of ET in layered MOFs.
\\~\\
\noindent\textbf{\large IV. Conclusions\label{Conclusions}}
\\~\\
The high degree of module ordering achievable within crystalline layered MOFs provides a basis for systematically relating structure and composition to photon capture and ET. 
In this work, we showed for the first time that one can model the through-space ET in layered MOFs by evaluating the geometric dependencies of the short and long range couplings within a  linear  array of SBUs along the stacking direction. 
By developing a Frenkel/CT Holstein Hamiltonian, we showed that photophysical properties of layered MOFs including their absorption spectra critically depend on the degree of ordering between layers. Findings in this direction not only advance our understanding of structure-transport-property relationships but can also help uncover new materials for nano and mesoscale optoelectronic applications.
\\~\\
\textbf{SUPPLEMENTARY MATERIAL}
Computational details and a comprehensive study on evaluating the effect of SBU's and OLU's structural changes on $\lambda$ are provided. Examples of input and output data for evaluating $\lambda$ are also provided.
\\~\\
\textbf{ACKNOWLEDGMENTS}
This work is supported by start-up fund from New Jersey Institute of Technology (NJIT) and used the Extreme Science and Engineering Discovery Environment (XSEDE), which is supported by National Science Foundation grant number CHE200007. This research has been partially enabled by the use of computing resources and technical supports provided by the HPC center at NJIT.
\\~\\
\textbf{AUTHOR DECLARATIONS}
\\~\\
\textbf{Conflict of Interest}
The authors have no conflicts of interest to declare.
\\~\\
\textbf{DATA AVAILABILITY}
The data that support the findings of this study are available in the paper and the supplementary material. Additional data are available from the corresponding authors upon reasonable request.

%\section*{References}
%%
%\bibliographystyle{aipnum4-1} 
\bibliography{bib}

\begin{thebibliography}{61}
\bibitem [{Noteb()}]{Noteb}%
  \BibitemOpen
  \bibinfo {note} {Here the assumption is that consecutive units are far enough
  apart that electron exchange and correlation contributions can be safely
  ignored.\cite {bitt09}}\BibitemShut {NoStop}%
\end{thebibliography}
%\bibliographystyle{aipnum4-1}

%\begin{tocentry}
%\includegraphics[scale=0.25]{toc1_EX}
%TOC entry
%\end{tocentry}


%\begin{thebibliography}{00}


%\bibitem{}

%\end{thebibliography}
\end{document}